# Comparative study of YSO, GaGG, and BGO scintillators coupled to a SiPM array for gamma-ray spectroscopy

S. Nuruyev [a,d], G. Ahmadov [a,b,c,d], D. Berikov [d,e], M. Holik [f,g], F. Ahmadov [a,b], A. Sadigov [a,b], R. Akbarov [a,b,d], J. Naghiyev [b], I. Nuruyev [a], A. Mammadli [a]

[a] *Institute of Radiation Problems- Ministry of Science and Education*
  *B.Vahabzade str.9, AZ1143, Baku, Azerbaijan*

[b] *Department of Nuclear Research of IDDA,*
  *Baku Shamakhy HW 20 km, Gobu sett. of Absheron dist., AZ 0100, Baku, Azerbaijan*

[c] *Khazar University*
  *41 Mahsati Str., AZ1096, Baku, Azerbaijan*

[d] *Joint Institute for Nuclear Research*
  *Joliot-Curie 6, 141980 Dubna, Russia*

[e] *The Institute of Nuclear Physics*
  *1 Ibragimova, 050032 Almaty, Kazakhstan*

[f] *Institute of Experimental and Applied Physics, CTU in Prague*
  *Horská 3a/22, 128 00 Praha 2, Prague, Czech Republic*

[g] *Faculty of Electrical Engineering, UWB in Pilsen*
  *Univerzitní 26, 306 14 Pilsen, Czech Republic*

*E-mail*: sebuhinuruyev@jinr.ru

ABSTRACT: A compact gamma-ray detector module was developed and characterized for spectroscopic applications across a wide energy range. The detector comprises a 4×4 array of silicon photomultipliers (SiPMs, Hamamatsu MPPC S13360-3050) coupled with three different inorganic scintillators (YSO, GaGG, BGO). The performance of the detector was evaluated over an energy range from 32 keV to 2500 keV using various standard gamma-ray point sources ($^{241}$Am, $^{137}$Cs, $^{22}$Na, $^{60}$Co, $^{228}$Th). All three detector configurations demonstrated linearity between pulse height and gamma-ray energy over most of the tested range. GAGG exhibited the most consistent linear behavior up to 2500 keV. YSO with a faster decay constant showed linearity up to ~2000 keV, followed by slight saturation effects likely due to SiPM microcell occupancy limits at higher photon flux. BGO, characterized by a low light output and longer decay time, demonstrated stable linearity up to 2500 keV but reduced sensitivity below 300 keV. The measured energy resolution at 662 keV was 8.17% for GAGG, 9.3% for YSO, and 10.2% for BGO, consistent with their light yield and photon detection efficiency (PDE) - matching characteristics. The obtained results showed the trade-offs between scintillator materials in terms of energy resolution, dynamic range, and linearity, and confirm the suitability of the tested configurations for compact gamma-ray spectroscopy applications across the studied energy range.

## Contents

1. **Introduction**

Scintillation detectors are essential instruments in the detection of ionizing radiation and have found widespread application in high-energy physics, medical imaging, astrophysics, and homeland security [1, 2]. These detectors operate by converting the energy of incident particles or photons such as alpha and beta particles, gamma rays, X-rays, or neutrons into visible photons within a scintillating medium. The emitted light is then registered by a photosensitive device, enabling the quantification and spectral analysis of the radiation. The choice of scintillator material, whether organic or inorganic, is determined by the specific requirements of the application, particularly in terms of energy resolution, timing, and detection efficiency. In gamma-ray spectroscopy, the energy range of interest spans from a few keV (e.g., for low-energy X-rays) to several MeV, relevant in applications such as nuclear activation analysis, radiological monitoring, and space-based gamma-ray observations [3, 4, 5]. Inorganic scintillators are typically preferred for such applications due to their high density, atomic number, and light yield, which together enable efficient detection and good energy resolution. Traditionally, photomultiplier tubes (PMTs) have been the standard readout technology for scintillation detectors due to their high gain and low noise. However, SiPMs have emerged as a compelling alternative. SiPMs offer numerous advantages, including compactness, low operating voltage, immunity to magnetic fields, mechanical robustness, radiation hardness and compatibility with monolithic integration [6, 7]. These features make them particularly attractive for the development of compact, portable, and power-efficient detection systems.

Despite the increasing adoption of SiPMs in gamma-ray detection, selecting an optimal scintillator material remains a nontrivial task. Key material properties, such as light yield, decay time, density, emission spectrum, and transparency, must be considered in conjunction with the photon detection efficiency (PDE) of SiPM, saturation behavior, and noise characteristics. The spectral match between the emission peak of scintillator and the PDE of SiPMs curve plays a critical role in determining overall system performance. While various SiPM–scintillator configurations have been explored in the literature, most studies are limited to a single scintillator type, a narrow energy range, or differing experimental conditions that preclude direct comparison [8-12]. Comprehensive, head-to-head evaluations of multiple scintillator materials under controlled and identical conditions remain scarce [13-17]. Yet such comparisons are essential to inform detector design for applications where compactness, resolution, and linearity must be balanced against cost and operational constraints. This study aims to address the current gap in comparative characterization of SiPM-based scintillation detectors and to provide practical insight into the selection of scintillator materials for compact gamma-ray spectroscopy systems [18]. The results also contribute to understanding of SiPM-scintillator interaction effects, including the impact of photon yield, decay time, and emission wavelength on detector performance.

## 2. Experimental setup and methodology

### 2.1 Detector module

The core component of the detection system is a compact 4×4 array of multi-pixel photon counters (MPPCs), specifically the Hamamatsu S13360-3050 model. Each SiPM element in the array features an active area of 3×3 mm², with a pixel density of 400 microcells/mm², and a photon detection efficiency (PDE) of approximately 40% at 470 nm well matched to the emission spectra of the scintillators used. The array was operated at a fixed bias voltage of 54.5 V, in accordance with the manufacturer's specifications for optimal gain and PDE [19]. The choice of this particular SiPM model is consistent with our group's prior investigations into photosensor performance for radiation detection applications [20-22].

To evaluate the effect of scintillator material on system performance, three different inorganic scintillators were tested: BGO, GAGG, and YSO [3]. All crystals were procured from Epic Crystal and machined to identical dimensions of 10×10×10 mm³ to ensure uniform optical coupling geometry. Their fundamental physical properties such as density, light yield, emission peak, and decay time are summarized in Table 1.

**Table 1.** Key properties of the investigated scintillator crystals.

| Crystal | BGO | GAGG | YSO |
|---|---|---|---|
| **Density** | 7,13 | 6,6 | 4,4 |
| **Light Output** | 8500 | 42000 | 24000 |
| **Wavelength of Emission Peak** | 480 | 520 | 410 |
| **Decay Constant** | 300 | 90 | 62 |

For optimal light collection, the scintillator surfaces (excluding the optical coupling face) were wrapped with multiple layers of white Teflon tape, serving as a diffuse reflector. The unwrapped bottom face of each crystal was optically coupled to the SiPM array using a transparent optical coupling gel, to minimize photon loss at the interface and ensure consistent optical transmission [2, 3].

**2.2 Data acquisition and high voltage system**

The experimental setup used for detector characterization is schematically illustrated in figure 1. The detector module was placed inside a custom-designed light-tight enclosure to eliminate ambient light interference. Gamma-ray sources were positioned above the crystal for irradiation under controlled conditions.

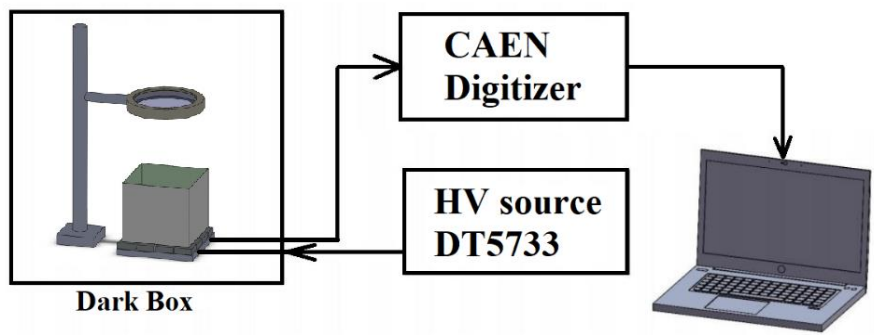

**Figure 1.** Schematic of the experimental setup used for detector characterization. The SiPM–scintillator module is enclosed in a dark box. Signal readout is handled by a CAEN DT5720B digitizer, while the SiPM bias is supplied by a CAEN DT5733 high-voltage source. Data are transferred to a PC for real-time monitoring and offline analysis.

The analog output signals from the SiPM array were processed using a CAEN DT5720B waveform digitizer, featuring a 12-bit resolution and a sampling rate of 250 MS/s. Data acquisition was performed in self-trigger mode, allowing autonomous event recording based on threshold crossings. The SiPM array was biased through a CAEN DT5733 high-voltage power supply, configured via USB control. The digitized waveforms were transferred in real time to a host PC for storage, waveform visualization, and offline signal processing. All data acquisition and monitoring tasks were controlled through custom scripts and CAEN-provided software [1-3, 18].

**2.3 Measurement procedure**

All measurements were carried out at a controlled ambient temperature of 22 °C to ensure thermal stability of the SiPM array. The detector module was irradiated with gamma-rays from standardized calibration sources: $^{241}$Am, $^{137}$Cs, $^{22}$Na, $^{60}$Co and $^{228}$Th), providing photon energies ranging approximately from 32 keV to 2500 keV. Each acquisition run was performed with an integration time of 5 minutes to ensure sufficient statistical accuracy. The charge integration gate for the digitizer was set according to the scintillator's decay constant: 400 ns for YSO and GaGG, and 750 ns for BGO.

Data analysis was performed using a custom-developed script implemented within the ROOT data analysis framework (CERN). All fits were performed using the fitting tools provided within the ROOT framework, and convergence was verified by visual inspection and by minimizing the reduced chi-squared ($\chi^2$/ndf) of the fit.

## 3. Results and discussion

### 3.1 Energy resolution

Energy resolution is a critical performance metric for gamma-ray spectrometers, as it quantifies the detector's ability to distinguish between closely spaced energy lines. In this study, energy resolution was assessed for each scintillator-SiPM combination using the prominent 662 keV gamma-ray emission from $^{137}$Cs source. The resolution was evaluated by fitting the photopeak with a Gaussian function and computing the full width at half maximum (FWHM), expressed as a percentage of the peak centroid position (FWHM/channel × 100%) [2, 5, 6].

Figure 2 shows the gamma-ray spectra from $^{137}$Cs source (662 keV) measured with the three scintillator-SiPM combinations. The energy resolution at 662 keV was found to be 8.17% for GaGG, 9.3% for YSO, and 10.2% for BGO.

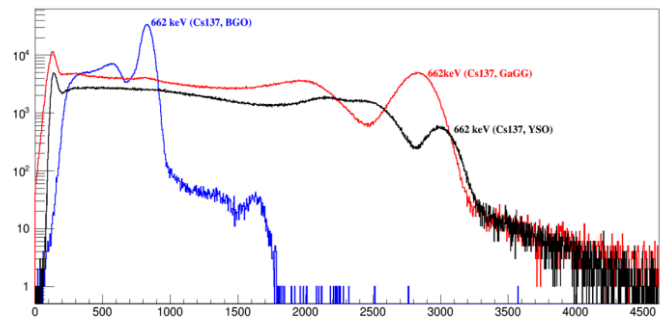

**Figure 2.** Gamma-ray spectra from $^{137}$Cs source (662 keV) measured with BGO, GAGG, and YSO scintillators coupled to a 4×4 SiPM array. The spectra were acquired under identical conditions, and energy resolution was extracted from Gaussian fits to the photopeaks.

The good energy resolution of GaGG is directly attributable to its very high intrinsic light yield (42000 ph/MeV), which produces more photoelectrons per event and reduces statistical fluctuations. YSO, although exhibiting a faster decay time and good spectral match with the SiPM PDE, provides a moderate light yield (~24000 ph/MeV), resulting in slightly poorer resolution. BGO demonstrated the lowest resolution, consistent with its low light yield (~8500 ph/MeV) and long decay constant (~300 ns), both of which degrade signal-to-noise characteristics and temporal integration efficiency. Furthermore, the 32 keV X-ray escape peak from $^{137}$Cs was not distinctly resolvable with the BGO detector, likely due to its long decay time and lower light yield in the low-energy region, a phenomenon also observed in other studies with low-light-yield scintillators [3].

### 3.2 Energy linearity

The linearity of each detector configuration was evaluated by analyzing the centroid positions of photopeaks across a set of gamma-ray sources with well-known energies. The measured channel number (proportional to pulse amplitude or integrated charge) was plotted as a function of the corresponding gamma-ray energy. Figures 3, 4, and 5 (left) present these calibration curves for the GAGG, YSO, and BGO scintillators, respectively. Each data point represents a photopeak fitted using a Gaussian function, with vertical error bars indicating uncertainty in peak position extraction.

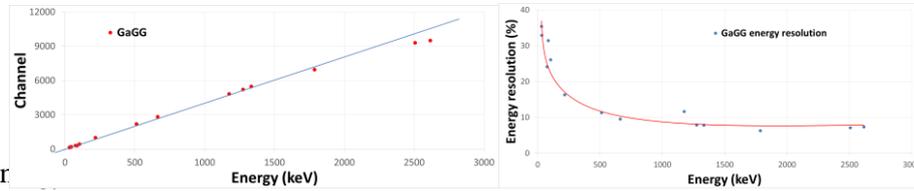

**Figure 3.** Energy calibration and resolution characteristics of the GaGG scintillator coupled to a 4×4 SiPM array. The left plot shows the energy calibration curve (channel number vs gamma-ray energy). The right plot shows energy resolution (FWHM%) as a function of energy.

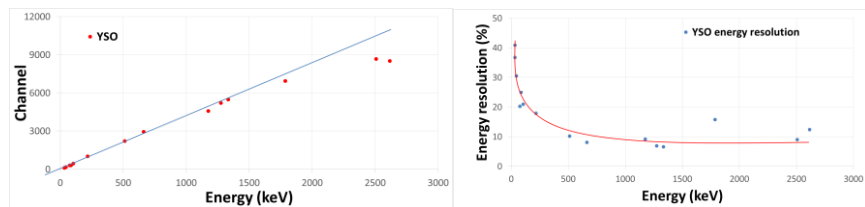

**Figure 4.** Energy calibration and resolution characteristics of the YSO scintillator. The left plot displays the linearity of detector response over the energy range. The right plot shows the dependence of energy resolution on gamma-ray energy.

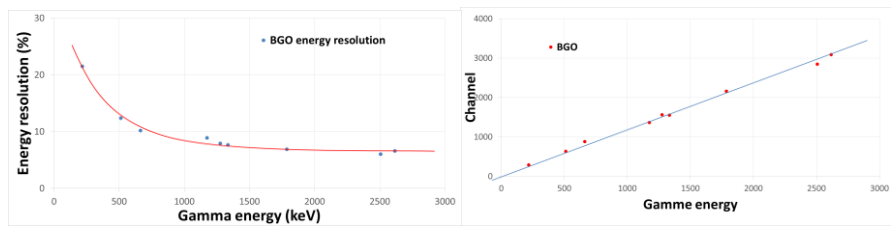

**Figure 5.** Energy calibration and resolution characteristics of the BGO scintillator. The left plot confirms a linear response across the full measurement range. The right plot shows the energy resolution (FWHM in %) as a function of gamma-ray energy.

Both GAGG and YSO exhibited an excellent linear response across the energy range from approximately 32 keV to 1332 keV, as indicated by the strong agreement between experimental data points and the linear fit. The high degree of linearity confirms the uniform charge output of these scintillators and their compatibility with the dynamic range of SiPM in this energy region.

In contrast, the BGO detector showed reduced sensitivity in the low-energy region (below ~300 keV), which is consistent with its lower light yield and longer scintillation decay time. These factors contribute to degrade signal amplitude and broadened peaks at low energies, making it difficult to resolve features such as the 122 keV line from $^{57}$Co or the 32 keV escape peaks from $^{137}$Cs. Nevertheless, BGO maintained a linear response over the entire measured range and showed no signs of saturation up to 2500 keV, which contrasts with the slight deviations from linearity observed at higher energies for the GAGG and YSO detectors. This behavior makes BGO a favorable choice for high-energy gamma-ray detection, as previously reported in studies of dense, non-hygroscopic scintillators [3].

As expected, the energy resolution (FWHM in %) improved with increasing gamma-ray energy across all scintillators, due to reduced relative statistical fluctuations in the number of detected photons. This trend is shown in the right-hand plots of Figures 3–5, which display the dependence of energy resolution on gamma energy for each scintillator.

## 4. Conclusion

In this study, a compact gamma-ray detector module based on a 4×4 silicon photomultiplier (SiPM) array was developed and characterized using three different inorganic scintillators (GAGG, YSO, and BGO). The system demonstrated good energy linearity and operational stability across a wide energy range, from 32 keV to 2500 keV, confirming the viability of SiPM-based detectors for various gamma-ray applications.

The results presented the performance trade-offs governed by the intrinsic properties of each scintillator. GAGG and YSO exhibited better energy resolution at low and medium energies, attributed to their higher light yields and fast decay times. These materials are well-suited for applications requiring compact detectors with high sensitivity and resolution in the sub-MeV to ~2000 keV energy range coupling the studied 4×4 SiPM array. In contrast, BGO, despite its lower light output and poorer resolution at low energies, offered clear advantages at higher energies, including enhanced stopping power, extended linear response, and resistance to saturation, making it a suitable choice for high-energy gamma-ray detection.

Overall, the detector modules developed in this work demonstrate potential for use in portable spectrometers, field-deployable radiation monitoring systems, and compact medical imaging or diagnostic devices.

## Acknowledgments

This work was supported by the Azerbaijan Science Foundation under grant agreement No [AEF-MGC-2024-2(50)-16/03/1-M-03], This work has received funding from the European Union's Horizon 2021 Research and Innovation Programme under the Marie Sklodowska-Curie's INNMEDSCAN project (grant agreement ID 101086178) and from the Science Committee of the Ministry of Science and Higher Education of the Republic of Kazakhstan (Grant no.BR20280986).